\documentstyle[12pt,eqsecnum,epsf,preprint,prd,aps]{revtex}
\begin{document}
\draft
\preprint{FTUAM 94--34}
\date{November 1994}
\title{
The $l=1$ Hyperfine Splitting in Bottomium\\
as a Precise Probe of the QCD Vacuum.\footnote{
This work is partially supported by CICYT, Spain.}}
\author{S.~Titard
\footnote{Electronic address:
{\tt stephan@nantes.ft.uam.es.}}
 \,and\,F.~J.~Yndur\'{a}in
}
\address{Departamento de F\'{\i}sica Te\'orica C-XI\\
Universidad Aut\'onoma de Madrid\\
28049 Madrid, Spain}
\maketitle
\begin{abstract}
By relating fine and hyperfine spittings for $l=1$ states in
bottomium we can factor out the less tractable part of the
perturbative and nonperturbative effects. Reliable predictions
for one of the fine splittings and the hyperfine splitting
can then be made calculating in terms of the remaining fine
splitting, which is then taken from experiment; perturbative
and nonperturbative corrections to these relations are under
full control. The method (which produces reasonable results
even for the $c{\bar c}$ system) predicts a value of 1.5 MeV
for the $(s\!=\!1)-(s\!=\!0)$ splitting in $b{\bar b}$, opposite in
sign to that in $c{\bar c}$. For this result the
contribution of the gluon condensate
$\langle \alpha_s G^2 \rangle$ is essential, as any model
(in particular potential models) which neglects this would give
a negative $b{\bar b}$ hyperfine splitting.
\end{abstract}

\section{Introduction}

It has been known for a long time that the short distance strong
interactions may be described by QCD in perturbation theory, and
that the {\em leading}, short distance nonperturbative effects can
be incorporated by taking into account the nonzero values of quark
and gluon condensates in the physical vacuum, $\vert {\rm vac}
\rangle$:
\begin{eqnarray}
\nonumber
\langle {\bar q} q \rangle
&\equiv&
\langle {\rm vac}\vert
:{\bar q}(0) q(0):
\vert {\rm vac} \rangle \neq 0 \;,
\\
\nonumber
\langle \alpha_s G^2 \rangle
&\equiv&
\langle {\rm vac}\vert
:G_{a\mu\nu}(0) G_a^{\mu\nu}(0):
\vert {\rm vac} \rangle \neq 0 \;.
\end{eqnarray}
{}From the pioneering SVZ work on QCD sum rules~\cite{bb:shif}, we
know that
\begin{equation}
\label{gluoncond}
\langle \alpha_s G^2 \rangle
= 0.042 \pm 0.020 \quad {\rm GeV^4}\;,
\end{equation}
a value confirmed by subsequent analyses. These methods are
applicable to study bound states of heavy quarks, as shown in the
papers of Leutwyler~\cite{bb:leut} and
Voloshin~\cite{bb:volo} and, more recently, in our
work~\cite{bb:ty,bb:tyii} where we have demonstrated that, indeed,
a consistent description of
$n=1$ $c{\bar c}$ states and $n=1,\,2$ $b{\bar b}$ ones
($n$ being the principal quantum number) is
obtained if one includes perturbative corrections in the form
of radiative corrections to the Coulombic, short
distance QCD potential,
\begin{equation}
\nonumber
\frac{- C_F \alpha_s}{r}\;,\; C_F= 4/3
\end{equation}
as well as nonperturbative ones through the
contributions of
$\langle \alpha_s G^2 \rangle$
(the quark condensate contributes a negligible amount).
In particular in Ref.~\cite{bb:tyii}
we found the following values for the
fine and hyperfine splittings in bottomium, with $n=2,\,l=1$,
$s$ the total spin, and $j$ the total angular momentum:
\begin{eqnarray}
\nonumber
\Delta_{21}&\equiv&
M(\chi_b(j=2))
-
M(\chi_b(j=1))
= 21
\phantom{a}^{+3}_{-6} \mp 2 \;{\rm MeV}
\\
\label{deltas}
\Delta_{10}&\equiv&
M(\chi_b(j=1))
-
M(\chi_b(j=0))
= 29
\phantom{a}^{+5}_{-9} \mp 3 \;{\rm MeV}
\\
\label{hyperf}
\vspace{0.8cm}
\Delta_{\rm hf}&\equiv&
M_{\rm average}(\chi_b(s=1,l=1))
-
M(\chi_b(s=0,l=1))
= 1.5  \mp 0.5 \pm 0.5 \;{\rm MeV}\;.
\end{eqnarray}
$\Delta_{21}$ and $\Delta_{10}$ may be compared to the experimental figures
\begin{equation}
\label{expt}
\Delta_{21}^{\rm exp}= 21 \pm 1 \;{\rm MeV},
\,\;
\Delta_{10}^{\rm exp}= 32 \pm 2 \;{\rm MeV}\,.
\end{equation}
In Eqs.~(\ref{deltas}) and (\ref{hyperf}) the first error is
due to that in the QCD parameter $\Lambda$,
\begin{equation}
\label{lambda}
\Lambda(4\;{\rm flavours},{\rm 2\, loops}) = 200 \phantom{a}^{+80}_{-60}
\, {\rm MeV} \;,
\end{equation}
and the the second to that of
$\langle \alpha_s G^2 \rangle$
as given by Eq.~(\ref{gluoncond}).

It is remarkable that the prediction of
$\Delta_{\rm HF}$ for $b{\bar b}$
suggests a value {\em opposite} in sign to that
of ${\bar c}c$ (where experimentally,
$\Delta_{\rm HF}^{\rm exp}({\bar c}c)= -0.9\;{\rm MeV}$).
This change of sign
is due to the structure of the QCD vacuum through the
contribution of the gluon condensate. In fact, and as we will
show below, any calculation neglecting this would give
a {\em negative}
$\Delta_{\rm HF}$, of the order of $-1$ to $-2$ MeV. Thus,
a measurement of $\Delta_{\rm HF}$
can be directly interpreted as a measurement of
$\langle \alpha_s G^2 \rangle$.

The results reported above,
Eqs.~(\ref{deltas}) and (\ref{hyperf})
are less impressive than what the agreement with experiment
would lead one to believe. The reason is that they contain
radiative and nonperturbative contributions
which are of relative order unity, thus impairing the
reliability of the calculation. In this note, however, we
show that by {\em combining} the fine (Eq.~(\ref{deltas})) and hyperfine
(Eq.~(\ref{hyperf})) splittings one can get a clean prediction for the
last, in which both radiative and nonperturbative effects
are small and fully under control.

\section{Radiative and nonperturbative interactions.}

As shown by several people (cf. Refs~\cite{bb:ty,bb:tyii} for details
and references) the leading perturbative, radiative and nonperturbative
interactions that contribute to the fine and hyperfine splittings are
the LS, T (tensor) and HF (hyperfine) potentials,
\begin{eqnarray}
\label{eq:hls}
V_{\rm LS}(\vec r) &=& \frac{3 C_F \alpha_s(\mu^2)}{2 m^2 r^3}
 \,\vec L \cdot \vec S
\,\left\{ 1 + \left[
 \frac{\beta_0}{2}(\ln r \mu - 1) +2 (1-\ln m r)
 +\frac{125 - 10\,n_f}{36} \right] \frac{\alpha_s}{\pi} \right\}
\\
\nonumber
&\equiv&
V_{\rm LS}^{(0)}
+ V_{\rm LS}^{({\rm rad})}
\;,\;
V_{\rm LS}^{(0)}
= \frac{3 C_F \alpha_s(\mu^2)}{2 m^2 r^3}
 \,\vec L \cdot \vec S  \;,
\\
V_{\rm T}(\vec r)
&=& \frac{C_F \alpha_s(\mu^2)}{4 m^2 r^3}
\label{eq:ht}
\,
\left\{
 1 + \left[
 D +\frac{\beta_0}{2} \ln r \mu - 3\ln m r \right]
  \frac{\alpha_s}{\pi}  \right\}
\\
\nonumber
&\equiv&
V_{\rm T}^{(0)}
+ V_{\rm T}^{({\rm rad})}
\;,\;
V_{\rm T}^{(0)}
= \frac{C_F \alpha_s(\mu^2)}{4 m^2 r^3} \;,
\\
\label{eq:hhf}
V_{\rm HF}(\vec r)
&=&
\left(\frac{\beta_0}{2}-\frac{21}{4}\right)\,
\frac{C_F \alpha_s(\mu^2)}{3 m^2\,r^3} \,{\vec S}^2 \;;
\end{eqnarray}
with
\begin{eqnarray}
\nonumber
&& S_{12}(\vec r) = 2\,\sum_{i j}S_i S_j \,
\left(\frac{3}{r^2} r_i r_j - \delta_{ij} \right)
\,,
\\
\nonumber
&& {\vec S}= {\vec S}_1 + {\vec S_2}
\,,\; {\vec L}= {\vec r} \times {\vec p}\,,
\\
\nonumber
 && C_A = 3, \, T_F = 1/2, \,\beta_0 = 11 - \frac{2\, n_f}{3},
\\
\nonumber
 && D = \frac{4}{3} \left(3 - \frac{\beta_0}{2} \right)
 + \frac{65}{12} - \frac{5\, n_f}{18}.
\end{eqnarray}
$V_{\rm HF}$ has an extra piece
proportional to $\delta({\vec r})$
which however does not contribute
to the states with $l=1$ in which we are interested.
A spin independent radiative correction which also intervenes
indirectly is given by
\begin{equation}
V_{\rm SI}(r)=
 - \frac{C_F (a_1 +\gamma_{\rm E}\,\beta_0/2)}{\pi r}\alpha_s^2
- \frac{C_F \beta_0 \alpha_s^2\ln r \mu}{2\pi r}\,;
\; a_1= \frac{31\,C_A -20\,n_f T_F}{36}\,.
\end{equation}
The nonperturbative interactions are generated by a term
\begin{equation}
\label{nonp}
 - g  {\vec r} \cdot
{{\vec {\cal E}}\mkern -7mu{\lower1.6ex\hbox{$\widetilde {}$}}\;\,}
 + \frac{g}{2 m^2} ({\vec S}\times {\vec p})  \cdot
{{\vec {\cal E}}\mkern -7mu{\lower1.6ex\hbox{$\widetilde {}$}}\;\,}
 - \frac{g}{m} ({\vec S}_1 - {\vec S}_2)  \cdot
{{\vec {\cal B}}\mkern -7mu{\lower1.6ex\hbox{$\widetilde {}$}}\;\,}
 \,.
\end{equation}
The constant chromoelectric
${{\vec {\cal E}}\mkern -7mu{\lower1.6ex\hbox{$\widetilde {}$}}\;\,}$
and chromomagnetic
${{\vec {\cal B}}\mkern -7mu{\lower1.6ex\hbox{$\widetilde {}$}}\;\,}$
fields
are to be taken as matrices in color space, and the vacuum is to
be assumed such that
\begin{eqnarray}
\nonumber
&& \langle \vec {\cal E}\, \rangle \,=\,
 \langle \vec {\cal B} \, \rangle \,=\, 0 \;,
\\
\nonumber
&& \langle g^2 {\cal B}^i_a {\cal B}^j_b  \rangle =
 - \langle g^2 {\cal E}^i_a {\cal E}^j_b  \rangle =
\frac{\pi \delta_{ij} \delta_{ab}}{3 (N_c^2-1)}
 \langle \alpha_s G^2  \rangle
\end{eqnarray}
($a,\,b$ colour indices, $i,\,j$ spatial ones).

The key point in the present paper is the remark
that the radial operator
that appears in all three
$V_{\rm LS}$,
$V_{\rm T}$ and
$V_{\rm HF}$ is
the same one at leading order,
viz., $r^{-3}$; and it so happens that
the {\em largest} perturbative and nonperturbative corrections are
those to the {\em wave functions}
which are the same for all fine and
hyperfine splittings. This allows us to factor these out,
being left with {\em small} and manageable pieces.

\section{Fine and Hyperfine splittings.}

Consider for example the fine splittings. Because the radiative pieces
of $V_{\rm LS}$, $V_{\rm T}$ are to be taken in perturbation
theory, and the same is true of the nonperturbative interactions
given in Eq.~(\ref{nonp}), we find, e.g.,
\begin{eqnarray}
\nonumber
\Delta_{21} &=&
\langle \Psi_{j=2} \vert
\left( V_{\rm LS} + V_{\rm T} \right)
\vert \Psi_{j=2} \rangle
-
\langle \Psi_{j=1} \vert
\left( V_{\rm LS} + V_{\rm T} \right)
\vert \Psi_{j=1} \rangle
\\
\nonumber
&&\,
+ 2\,
\Bigl\langle \Psi_{j=2} \Bigr\vert
\left( - g  {\vec r} \cdot
{{\vec {\cal E}}\mkern -7mu{\lower1.6ex\hbox{$\widetilde {}$}}\;\,}
\right)
\,\frac{1}{H^{(0)}-E^{(0)} } \,
\frac{g}{2 m^2} ({\vec S}\times {\vec p})  \cdot
{{\vec {\cal E}}\mkern -7mu{\lower1.6ex\hbox{$\widetilde {}$}}\;\,}
\Bigl\vert \Psi_{j=2} \Bigr\rangle
\\
\label{expr}
&&\,
- 2\,
\Bigl\langle \Psi_{j=1} \Bigr\vert
\left( - g  {\vec r} \cdot
{{\vec {\cal E}}\mkern -7mu{\lower1.6ex\hbox{$\widetilde {}$}}\;\,}
\right)
\,\frac{1}{H^{(0)}-E^{(0)} } \,
\frac{g}{2 m^2} ({\vec S}\times {\vec p})  \cdot
{{\vec {\cal E}}\mkern -7mu{\lower1.6ex\hbox{$\widetilde {}$}}\;\,}
\Bigl\vert \Psi_{j=1} \Bigr\rangle \,.
\end{eqnarray}
The $\Psi_j$ are the wave functions for the states with
$n=2$, $l=1$, $s=1$ and total angular momentum $j$. The contributions
to $\Delta_{21}$ may be split into two pieces. First we have what we
may call {\em "external"}, $\Delta^{\rm ex}$, obtained by substituting
in Eq.~(\ref{expr}) the unperturbed wave functions solutions to
\begin{displaymath}
H^{(0)}
\Psi_j^{(0)}
= E^{(0)}
\Psi_j^{(0)} \;,
\end{displaymath}
where the potential in $H^{(0)}$ is just the Coulombic one. It so
happens that both {\em radiative} and {\em nonperturbative}
contributions to $\Delta^{\rm ex}$ are {\em small}, at the 10\% level
or smaller.

The troublesome piece is what we may call {\em "internal"},
$\Delta^{\rm in}$, and is due to the fact that $\Psi_j$ also contains
spin--independent radiative and nonperturbative corrections:
\begin{displaymath}
\Psi_j
=
\Psi_j^{(0)}
+
\Psi_j^{\rm rad}
+
\Psi_j^{\rm NP}\;.
\end{displaymath}
Then, $\Delta^{\rm in}$
would be the contribution of
$\Psi_j^{\rm rad}$, $\Psi_j^{\rm NP}$ to Eq.~(\ref{expr})
(the radiative corrections
are caused by the spin--independent corrections to the
potential, and the non perturbative ones
by the spin--independent pieces generated
by Eq.~(\ref{nonp}), i.e, the contribution quadratic in
$
 g  {\vec r} \cdot
{{\vec {\cal E}}\mkern -7mu{\lower1.6ex\hbox{$\widetilde {}$}}\;\,}
$).
As stated before, however, the key point is that, when
evaluating $\Delta^{\rm in}$, and because
$\Psi_j^{\rm rad}$ and $\Psi_j^{\rm NP}$ are already
perturbations, only the leading pieces of the potentials
i.e.,
$V^{(0)}_{\rm LS}$,
$V^{(0)}_{\rm T}$
and
$V_{\rm HF}$ have to be considered, and these are all proportional
to $r^{-3}$, hence identical for fine and hyperfine splittings.

For a precise evaluation we take the explicit formulas of
Ref.~\cite{bb:tyii}. Then one finds the following
theoretical values for the splittings:
\begin{eqnarray}
\nonumber
\Delta_{10}^{\rm th} &=&
\frac{5}{4} \left( 1 + \delta_{\rm rad} \right)
\Delta_{21}^{\rm th} \, - \delta_{\rm NP} \,,
\\
\nonumber
\delta_{\rm rad} &=&
\left[
 \frac{3}{4} \ln\frac{C_F \widetilde{\alpha}_s }{2}
+ \frac{80 + 13 n_f}{96} + \frac{3}{4} \gamma_{\rm E}
\right] \frac{\alpha_s}{\pi} \,,
\\
\label{theo}
\delta_{\rm NP} &=&
\frac{2244}{3315} \, \frac{\pi \langle \alpha_s G^2 \rangle}{m^3
(C_F \widetilde{\alpha}_s)^2} \,,
\\
\nonumber
\widetilde{\alpha}_s
&=&
\left[ 1 + \frac{\gamma_{\rm E} \beta_0/2 + (93- 10 n_f)/36}{\pi}
\alpha_s \right]
\,\alpha_s \,.
\end{eqnarray}
As for the hyperfine splitting
\begin{equation}
\label{theohf}
\Delta_{\rm HF}^{\rm th}=
\frac{5}{24}
\left(\frac{\beta_0}{2}-\frac{21}{4}\right)\,
C_F \alpha_s \Delta_{21}^{\rm th}
+\frac{976}{1053} \, \frac{\pi \langle \alpha_s G^2 \rangle}{m^3
(C_F \widetilde{\alpha}_s)^2} \,.
\end{equation}
The nonperturbative piece of Eq.~(\ref{theohf}) is due to
the term
$\displaystyle
 - \frac{g}{m} ({\vec S}_1 - {\vec S}_2)  \cdot
{{\vec {\cal B}}\mkern -7mu{\lower1.6ex\hbox{$\widetilde {}$}}\;\,}
$ in Eq.~(\ref{nonp}).
In Ref.~\cite{bb:tyii}, the best overall fit to $n=2$ states was
obtained for $\alpha_s(\mu^2)=0.38$ (this corresponds to $\mu$=0.93
GeV). In this paper we will allow
$\alpha_s(\mu^2)$ to vary between 0.33 and 0.43 which  corresponds
to the range $2\,{\rm GeV}\ge \mu \ge 0.8$ GeV, the expected "relevant"
scale, $\mu \sim \langle {\vec k}^2 \rangle_{21}^{1/2}$. In all this
range
$\vert \delta_{\rm rad} \vert
\,{<\mkern -19mu{\lower1.3ex\hbox{$\sim$}\;}\,}
10 \%$, and
$\vert \delta_{\rm NP} \vert
\,{<\mkern -19mu{\lower1.3ex\hbox{$\sim$}\;}\,}
5 \%$: we check that both radiative and nonperturbative corrections
to the fine splittings, Eq.~(\ref{theo}), are small. Agreement with
experiment is excellent in all the range. This is shown is Fig.~1 where
we plot the values of
$\Delta_{21}^{\rm th}$,
$\Delta_{10}^{\rm th}$ that follow from Eq.~(\ref{theo})
by treating, in
Eq.~(\ref{theo}), $\Delta_{21}^{\rm th}$ as a free parameter,
then fitting $\Delta_{21}^{\rm th}$, $\Delta_{10}^{\rm th}$
to experiment. For all
the range, agreement between the theoretical values $\Delta^{\rm th}$
and experimental ones (cf. Eq.~(\ref{expt})) is better than 10\% with respect
to central experimental values, and
agreement within experimental errors
is even obtained for $\alpha_s=0.43$.

\section{Discussion and Results.}

Allowing $\alpha_s= 0.38 \pm 0.05$, and the range of Eq.~(\ref{gluoncond})
for
$\langle \alpha_s G^2 \rangle $,
Eqs.~(\ref{theo}) and (\ref{theohf}) yield a very reliable
prediction for the hyperfine splitting (see Fig.~2):
\begin{equation}
\label{predhf}
\Delta_{\rm HF}= 1.5 \phantom{a}^{+0.8}_{-1.2}\pm 0.5 \;{\rm MeV}\,
\end{equation}
(the first error is due to the variation of $\alpha_s$, and the
second to the error in $\langle \alpha_s G^2 \rangle$).

A few words should be said on this. As Eq.~(\ref{theohf}) shows,
$\Delta_{\rm HF}$ is the sum of two terms, a perturbative and a
nonperturbative one. That the second one dominates is due to the fact that
the perturbative contribution itself is the {\em difference}
between two pieces, proportional respectively to $\beta_0/2$ and $21/4$,
each one large, but which cancel almost exactly:
for $n_f=4$, $\beta_0/4=4.17$, while $21/4=5.25$. And the
whole perturbative term is still smaller because the {\em tree level}
potential is proportional to $\delta({\vec r})$, hence gives zero for
$l=1$ states. Thus an effect potentially ${\cal O}(\beta_0/2) \sim 4$ is
actually of order $(\beta_0/2-21/4)\,\alpha_s/\pi \sim -0.13$. This
is very much suppressed and thus highlights by contrast the
nonperturbative contribution.

A remarkable feature of the splitting (\ref{predhf})
is that it {\em cannot}
be reproduced by the use of any of the phenomenological potentials
available on the market. Indeed, {\em any} model that
neglects the nonzero expectation value of
$
{{\vec {\cal B}}\mkern -7mu{\lower1.6ex\hbox{$\widetilde {}$}}\;\,}
$
in the QCD vacuum will
{\em necessarily} yield a {\em negative} $\Delta_{\rm HF}$.
In particular, if one pretends to simulate nonperturbative effects by use
of spin--independent potentials, then one has
Eqs.~(\ref{theo}) and (\ref{theohf}) replaced by:
\begin{eqnarray}
\nonumber
\Delta_{10}^{\rm pot} &=&
\frac{5}{4} \left( 1 + \delta_{\rm rad} \right)
\Delta_{21}^{\rm pot}
\\
\label{pot}
\Delta_{\rm HF}^{\rm pot}&=&
\frac{5}{24}
\left(\frac{\beta_0}{2}-\frac{21}{4}\right)\,
C_F \alpha_s \Delta_{21}^{\rm pot} \,.
\end{eqnarray}
A simple calculation, also for the range
$\alpha_s= 0.38 \pm 0.05$
gives good agreement for
$\Delta_{10}^{\rm pot}$,
$\Delta_{21}^{\rm pot}$ with experiment, but now
\begin{equation}
\label{pothf}
\Delta_{\rm HF}^{\rm pot} = -1.2 \phantom{a}^{+0.3}_{-0.4}
\;{\rm MeV}\,.
\end{equation}
The gap between Eq.~(\ref{pothf}) and Eq.~(\ref{predhf}) is
sufficiently large that a measurement, probably feasible
with b--factories, should be able to reveal it.

As a last comment, let us remind the reader that the analysis we have
carried is justified only at short distances. For $b{\bar b}$ with
$n=2\,,l=1$,
$\langle r \rangle_{21} \sim (1 \,\,{\rm GeV})^{-1}$. For
$t{\bar t}$ the situation is even more favourable, but the
measurement is of course impossible.
For $c{\bar c}$ we cannot carry a rigourous analysis
since we have
$\langle r \rangle_{21}^{c{\bar c}} \sim (0.5 \,\,{\rm GeV})^{-1}$.
However we may attempt a phenomenological calculation
which
mimicks the theoretical one done just before; using Eq.~(\ref{theo})
with the appropriate changes, i.e $m=m_c$, $n_f =3$ and replacing
$\alpha_s \rightarrow \alpha_c^{\rm eff}$, we get
perfect agreement with the experimental fine splittings for
$\alpha_c^{\rm eff}= 0.86$. The external nonperturbative term is
reasonably small and,
although $\alpha_c^{\rm eff}$ is large, one can still interpret
it as an
{\em effective} coupling into which are lumped internal corrections and
higher order perturbative ones. Using then the appropriate
modification of Eq.~(\ref{theohf})
\begin{equation}
\label{theohfc}
\Delta_{{\rm HF},c}=
\frac{5}{24}
\left(\frac{\beta_0}{2}-\frac{21}{4}\right)\,
C_F \alpha_c^{\rm eff} \Delta_{21}^{\rm th}
+\frac{976}{1053} \, \frac{\pi \langle \alpha_s G^2 \rangle}{m^3
(C_F \widetilde{\alpha}_c^{\rm eff})^2} \,,
\end{equation}
we get
\begin{equation}
\label{predhfc}
\Delta_{{\rm HF},c}= -2.5 \pm 2.5\;{\rm MeV}\,.
\end{equation}
(the error being that in
$\langle \alpha_s G^2 \rangle$).
It is indeed remarkable that the experimental value
$\Delta_{{\rm HF},c}^{\rm expt}= -0.9\;{\rm MeV}$ falls within
the range of the previous prediction.
Another noteworthy feature is the role played by the gluon condensate
in obtaining (\ref{theohfc}). If we had set
$\langle \alpha_s G^2 \rangle=0$ we would have obtained
\begin{displaymath}
\Delta_{{\rm HF},c}= -9.2\;{\rm MeV}
\end{displaymath}
widely off experiment.\footnote{
In phenomenological papers agreement of
$\Delta_{{\rm HF},c}$
with experiment is obtained at the cost of using
{\em different} values of $\alpha_s$ for fine and hyperfine splittings,
or extra phenomenological LS or T interactions, or both; see
for example Ref.~\cite{bb:halz} and work quoted there.}
This strongly
suggests that the system of Eqs.~(\ref{theo}) and (\ref{theohf})
which as we have just seen works reasonably well even for $c{\bar c}$
bound states,
can be trusted
to provide a reliable description
of the $n=2$, $l=1$ fine and
hyperfine splittings of bottomium.

\newpage
\underline{Figure Captions.}
\leftline{\phantom{aaaa}
Fig.~1.- Experimental and Theoretical
(from Eq.~(\ref{theo})) fine splittings.
}
\leftline{\phantom{aaaa}
Fig.~2.-
Hyperfine splitting.
}
\leftline{\phantom{aaaaaa}
continuous line: central value for
$\langle \alpha_s G^2 \rangle$}
\leftline{\phantom{aaaaaa}
shaded area: varying
$\langle \alpha_s G^2 \rangle$ between its bounds}
\leftline{\phantom{aaaaaa}
dotted line: neglecting gluon condensate.}

%
%

\newpage
\begin{figure}[b]
    \leavevmode
    \epsfverbosetrue
    \epsffile{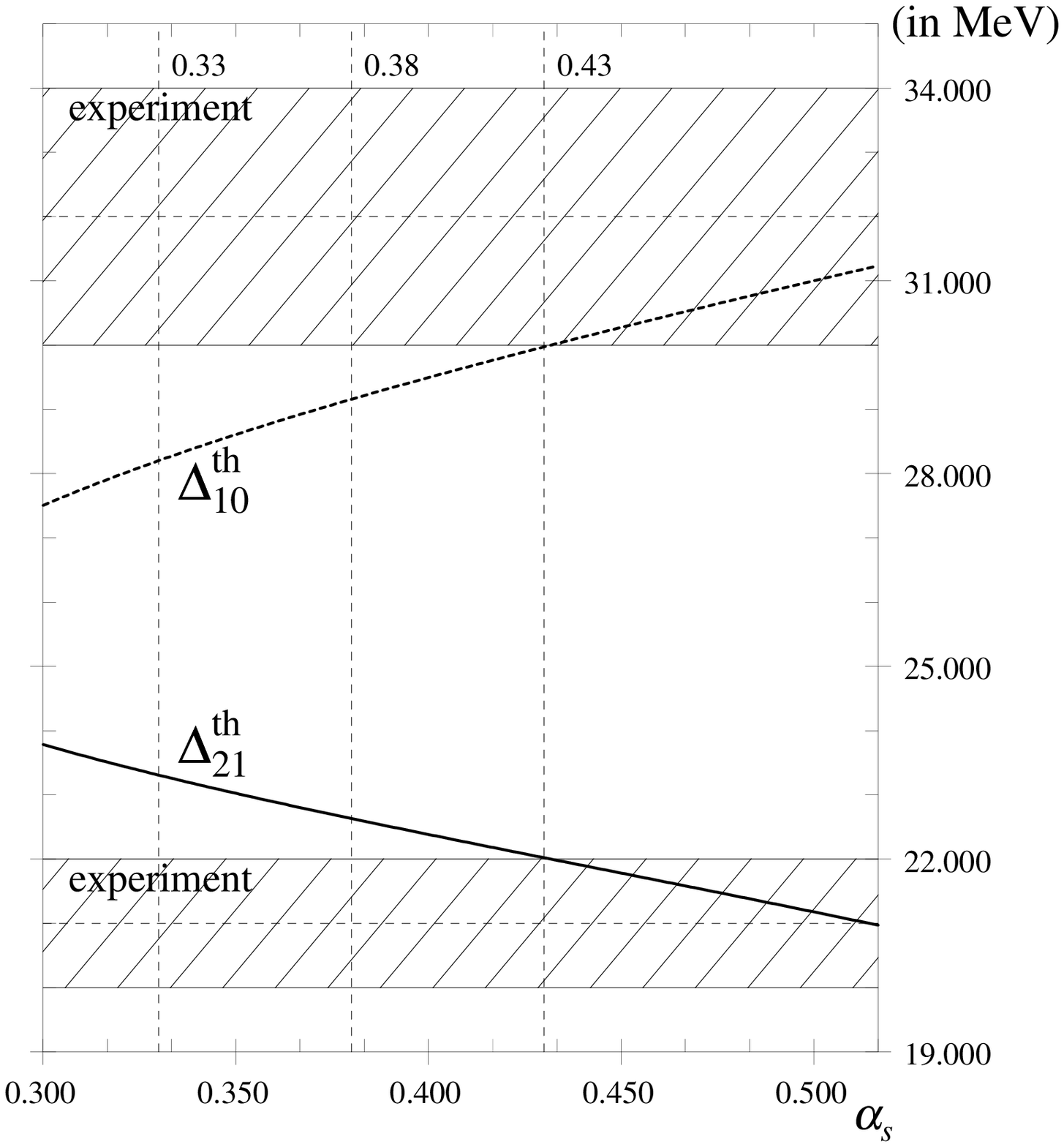}
\end{figure}
\centerline{Fig.~1}
\newpage
\begin{figure}[b]
    \leavevmode
    \epsfverbosetrue
    \epsffile{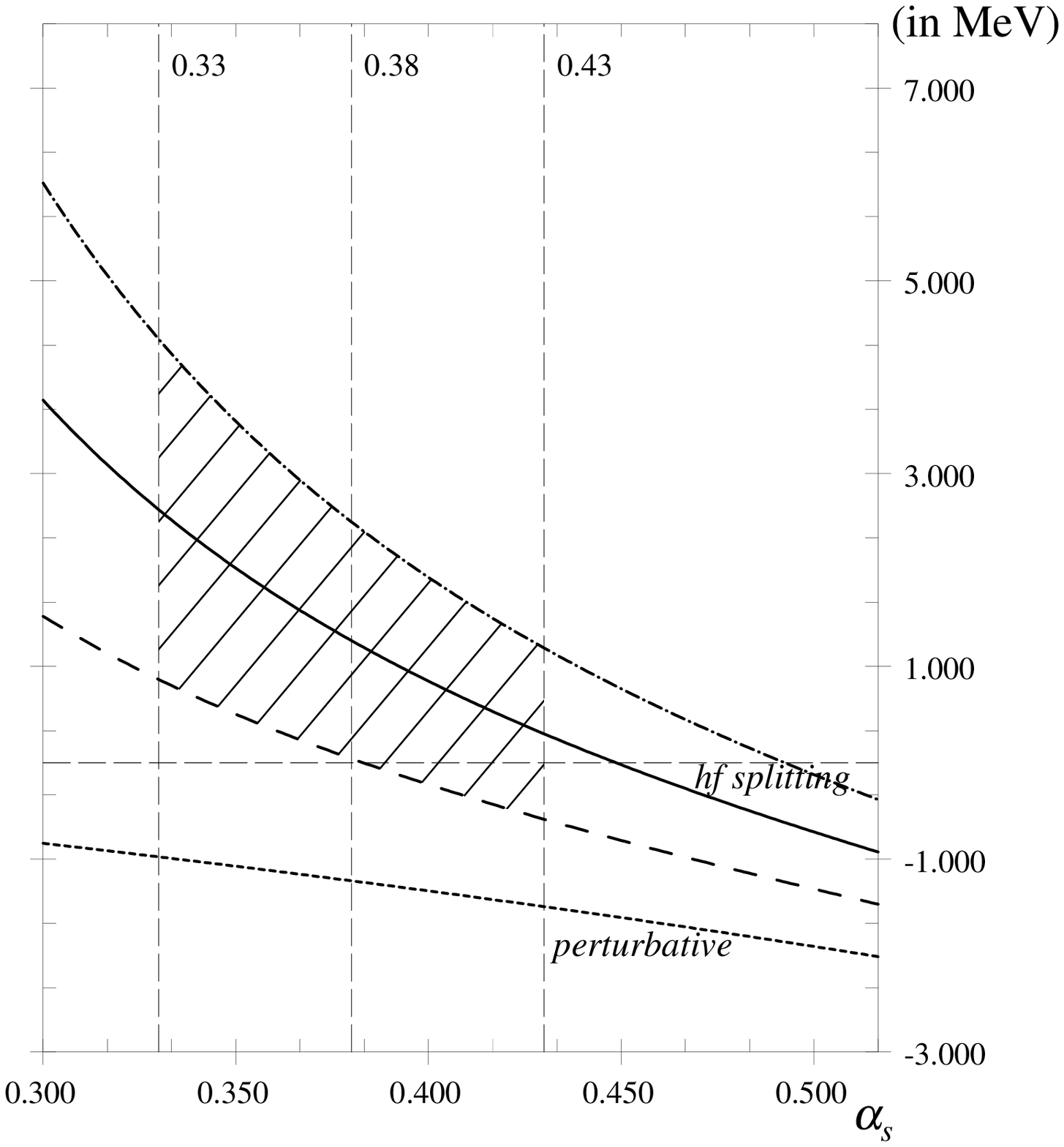}
\end{figure}
\centerline{Fig.~2}

\end{document}